%% file: main.tex
\documentclass[10pt, journal, letterpaper]{ IEEEtran}

\usepackage{graphicx}

\usepackage{todonotes} 
\let\oldtodo\todo
\renewcommand{\todo}[1]{\oldtodo[inline]{#1}}

\usepackage{color,soul} 

\usepackage[LGRgreek]{mathastext}
\usepackage{algorithm}
\usepackage{algorithmic}
\usepackage{amsmath}
\usepackage{booktabs}
\usepackage{comment}
\usepackage{stfloats}
\usepackage{booktabs}
\usepackage{multirow}
\usepackage{array}
\usepackage{subcaption}

\begin{document}

\input{myContents/title}
\input{myContents/authors}

\maketitle

\input{myContents/body}

\bibliographystyle{IEEEtran}
\bibliography{myContents/references}

\end{document}

%% file: myContents/title.tex
\title{ioPUF+: A PUF Based on I/O Pull-Up/Down Resistors for Secret Key Generation in IoT Nodes}

%% file: myContents/authors.tex
\author{
    Dilli Babu Porlapothula\IEEEauthorrefmark{1},
    Pralay Chakrabarty\IEEEauthorrefmark{1}, 
    Ananya Lakshmi Ravi \IEEEauthorrefmark{2}, \\
    and Kurian Polachan\IEEEauthorrefmark{1}, Senior Member, IEEE
    \\
    \IEEEauthorrefmark{1}International Institute of Information Technology, Bangalore, India\\
    \IEEEauthorrefmark{2}Indian Institute of Information Technology, Design and Manufacturing, Kancheepuram, India\\
    Email: \{dillibabu.porlapothula, kurian.polachan, pralay.chakrabarty\}@iiitb.ac.in, ec21b1004@iiitdm.ac.in

\thanks{An earlier version of this paper was presented at the IEEE 22nd Mediterranean Electrotechnical Conference (MELECON), 2024~\cite{ioPUF}}
}

%% file: myContents/body.tex
\input{myContents/section-1-abstract}

\input{myContents/section-2-introduction}

\input{myContents/section-3-related-work}

\input{myContents/section-4-system-design}

\input{myContents/section-5-results}

\input{myContents/section-5b-applications}
\input{myContents/section-6-conclusion}

%% file: myContents/section-1-abstract.tex
\begin{abstract}
In this work, we present ioPUF+, which incorporates a novel Physical Unclonable Function (PUF) that generates unique fingerprints for Integrated Circuits (ICs) and the IoT nodes encompassing them. The proposed PUF generates device-specific responses by measuring the pull-up and pull-down resistor values on the I/O pins of the ICs, which naturally vary across chips due to manufacturing-induced process variations. Since these resistors are already integrated into the I/O structures of most ICs, ioPUF+ requires no custom circuitry and no new IC fabrication. This makes ioPUF+  suitable for cost-sensitive embedded systems built from Commercial Off-The-Shelf (COTS) components.
Beyond introducing a new PUF, ioPUF+ includes a complete datapath for converting raw PUF responses into cryptographically usable secret keys using BCH error correction and SHA-256 hashing. Furthermore, ioPUF+ demonstrates a practical use case of PUF-derived secret keys in securing device-to-device communication using AES encryption.

We implemented ioPUF+ on the Infineon PSoC-5 microcontroller and evaluated its performance across 32 devices using standard PUF metrics. The results show excellent reliability of 100\% (zero intra-device Hamming distance), strong uniqueness (inter-device Hamming distance of 50.33\%), near-ideal uniformity (50.54\%), and negligible bit aliasing. Stability tests under temperature and supply-voltage variations show worst-case bit-error rates of only 2.63\% and 2.10\%, respectively. We also profiled the resource and energy usage of the complete ioPUF+ system, including the PUF primitive, BCH decoding, SHA-256 hashing, and AES encryption. The full implementation requires only 19.8 KB of Flash, exhibits a latency of 600 ms, and consumes 79 mW of power, demonstrating the suitability of ioPUF+ for resource-constrained IoT nodes.
\end{abstract}

\begin{IEEEkeywords}
Physical Unclonable Functions (PUFs), Internet of Things (IoT), hardware security
\end{IEEEkeywords}

%% file: myContents/section-2-introduction.tex
\section{Introduction}

\begin{figure*}[!htbp]
\centering
\includegraphics[width=0.90\linewidth]{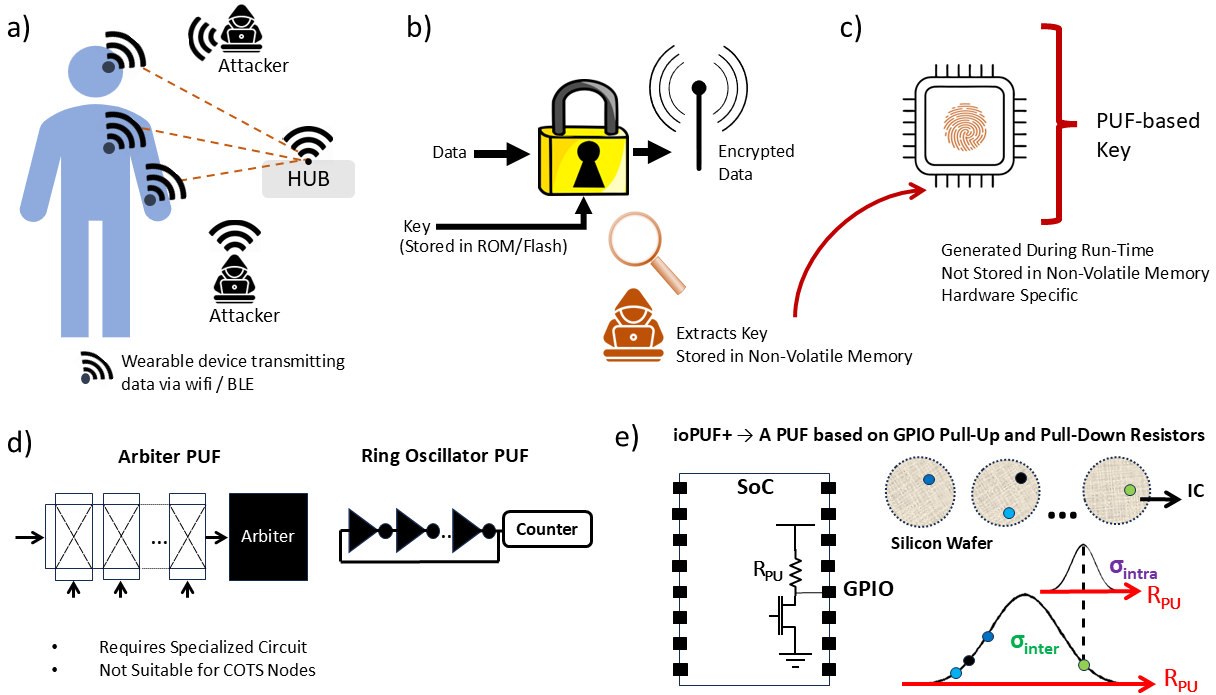}
\caption{(a) Security threat in IoT applications involving unencrypted data transfers over a wireless medium; eavesdropping attacks by nearby hackers.
(b) Conventional approach to safeguard against eavesdropping attacks; encrypted data transfers with the key used for encryption stored in the device’s non-volatile memory. Drawback → the key can be recovered via physical attacks on the device.
(c) PUF-based solution, where the keys used for encrypting data transfers are generated during device runtime. Keys are not stored in the device’s non-volatile memory; this resists physical recovery of the key.
(d) Existing types of PUFs require specialized circuits.
(e) Our proposed ioPUF+, based on I/O pull-up/down resistors.}
\label{iot_security_overview}
\end{figure*}

In many Internet of Things (IoT) applications, securing data is paramount. Consider the case of wearable sensors that continuously collect and transmit sensitive health information from a patient to a nearby hub over the air. If the transmitted data is not adequately protected, an adversary in the vicinity may eavesdrop on the wireless channel as illustrated in Figure~\ref{iot_security_overview}(a), gaining access to private health information, and potentially misuse it~\cite{iot_threat_meneghello}.

A similar risk exists when an IoT device stores sensitive information in its on-board non-volatile memory, such as an SD card. For example, an access-control system may locally store user identifiers and authentication logs. If an attacker gains physical access to the device, this data can be extracted~\cite{iot_threat_meneghello}. Such leaks are unacceptable in environments where personal or organizational information must remain confidential.

A common approach to secure data in transit or at rest is to encrypt it using a secret key, often stored in non-volatile memory of the device, following well-established standards such as AES. However, this method has a critical limitation: if an attacker gains physical access to the device  as shown in Figure~\ref{iot_security_overview}(b), the stored key can be extracted~\cite{security_iot_Abomhara}. Once compromised, the adversary can  eavesdrop on future encrypted communications or decrypt any stored information.

Public-key cryptography offers a solution, in particular for protecting data during transit. Here, the recipient’s public key is provisioned to the IoT node during an enrollment phase. Later, the node encrypts its data using the public key before transmitting it over the air. Only the intended recipient, who owns the corresponding private key, can decrypt it. However, this method has two major drawbacks. First, public-key cryptography incurs significantly higher computational and energy overheads compared to symmetric-key methods like AES, making it less suitable for resource and energy-constrained IoT nodes~\cite{low_power_iot}. Second, if an attacker learns the data format, they may create a malicious node that encrypts forged data using the same public key. The receiver may then process this malicious data as legitimate, leading to data-integrity violations~\cite{security_iot_Abomhara}. In safety-critical systems, such as in industrial IoT systems, such violations may trigger catastrophic failures or prolonged downtime.

Another possibility is to protect unauthorized readouts from the device's memory by storing the key. The method works by disabling the readout circuit to the memory once the key is logged in, so that external devices, e.g., programmers, cannot perform any memory reads from the device.  However, the method cannot prevent sophisticated invasive attacks, such as the case where the chip is decapsulated, and the internal circuitry is probed to infer the key.

Techniques can also be employed to detect physical intrusions or probing attempts to safeguard keys. However, such mechanisms are effective only when the device is powered on. Keys stored in volatile memory can be protected, but if the key resides in non-volatile memory (e.g., Flash, ROM, or SD cards), often that is the case, the device remains vulnerable when in powered-down mode~\cite{security_iot_mosenai}. 

As an alternative to the aforementioned traditional approaches, Physically Unclonable Function (PUF)-based methods for securing IoT nodes have been proposed. PUF-based techniques eliminate the need to store keys in non-volatile memory~\cite{puf_tutorial,puf_review_paper,gebali2022review}, thereby reducing the risk of key extraction and the overhead of protecting stored keys. Instead, they generate secret keys during the runtime by exploiting intrinsic hardware characteristics unique to each device, making these keys extremely difficult to duplicate as depicted in Figure~\ref{iot_security_overview}(c). The generated key resides only in volatile memory (e.g., RAM). As such, when the device powers off, the key disappears from the memory, preventing an attacker from extracting it through physical access. Furthermore, because the key exists only during the powered-on state, additional layers of protection, such as active tamper-detection sensors or probe detection circuits, which operate only when the device is powered, can be employed.

At the core of any PUF-based method for generating secret keys lies a PUF primitive, a circuit and/or algorithm that produces unique, device-specific responses that serve as a fingerprint of the hardware. Several types of PUF primitives have been reported in the literature. However, many of them suffer from a significant drawback: they require custom circuitry and, consequently, a dedicated Integrated Circuit (IC)~\cite{vaidya2020gpio}. For example, the Ring Oscillator PUF (RO-PUF)~\cite{RO-PUF} relies on a specially engineered ring-oscillator network. As such, many of the existing PUF primitives~\cite{A-puf,ReRam_puf, delay_puf_review} have limited application in IoT nodes, which are typically built using low-cost, resource-constrained Commercial Off-The-Shelf (COTS) components as illustrated in Figure~\ref{iot_security_overview}(d). Memory-based PUFs, such as SRAM PUFs~\cite{SRAM_PUF2}, are compatible with COTS microcontrollers but introduce their own constraints: response generation is typically restricted to the device’s boot phase. As a result, the derived secret key must be stored in memory throughout the device’s runtime, which may raise security  concerns.

In response to these limitations, we introduce ioPUF+, which incorporates a novel PUF primitive that generates unique fingerprints for Integrated Circuits (ICs), specifically, microcontroller ICs commonly used in IoT nodes, by measuring the values of the pull-up and pull-down resistors connected to the IC's input-output (I/O) pins as demonstrated in Figure~\ref{iot_security_overview}(e). Our approach is based on the observation that, in many IC designs, the pull-up/down resistor values exhibit substantial inter-device variation while showing much smaller intra-device variation across repeated measurements. By leveraging these resistor values, the proposed PUF primitive produces device-specific responses without requiring custom circuitry or new IC fabrication, as pull-up and pull-down resistors are already integral to the IO structures of most COTS microcontroller ICs. Moreover, unlike SRAM PUFs, which restrict response generation to the boot phase, the responses here can be generated at any point during device operation. Consequently, ioPUF+ based secret keys can be derived on a need basis and can be securely erased from memory after use, improving the security of the keys.

In this work, we present the design and implementation details of our proposed PUF primitive, including techniques for reliably measuring the pull-up/pull-down resistor values on IC I/Os and the associated algorithm for generating responses from these measurements. We also perform a detailed characterization of these responses across several devices to evaluate the quality of the primitive using standard PUF metrics such as reliability, stability, uniformity, and uniqueness. Beyond introducing a new PUF design, ioPUF+ also details the methods and implementation steps required to convert raw PUF responses into cryptographically usable secret keys. Finally, we demonstrate a practical use of our work, securing device-to-device communication, where data transfers between the devices are protected using AES encryption with ioPUF+ derived secret keys.

The rest of the paper is organized as follows: Section~\ref{sec_related_work} reviews related work on I/O-based PUFs and PUFs for secret key generation used in cryptography applications. Section~\ref{sec_sysdesign} presents the system design of our ioPUF+, the algorithm used for generating ioPUF+ responses, and the procedure for obtaining the PUF ID and secret key from the raw ioPUF+ response. Section~\ref{sec_evaluation} describes the experimental setup, including hardware implementation on the PSoC-5 microcontroller, evaluation of standard PUF metrics, and discussion of ioPUF+ results for these metrics. In Section~\ref{sec_applications}, we demonstrate cryptographic key generation, provide an end-to-end secure communication application, and discuss resource utilization for all components on the PSoC-5 microcontroller. Section~\ref{sec_conclusion} concludes the paper with future directions.

%% file: myContents/section-3-related-work.tex
\section{Related Work}
\label{sec_related_work}

ioPUF+ presented in this work is an extension of our recent study, ioPUF~\cite{ioPUF}. ioPUF was among the first works to conduct detailed simulation and experimental analyses to investigate the feasibility of using the internal pull-up and pull-down resistors of I/O pins for generating PUF responses in commercial off-the-shelf (COTS) microcontrollers.

However, the ioPUF~\cite{ioPUF} design had several shortcomings. First, it measured either the pull-up or the pull-down resistor of each I/O pin, limiting the PUF response length to 45 bits when using 10 I/O pins, an insufficient length for many cryptographic applications. Second, the PUF evaluation metrics, particularly the inter-device Hamming distance, fell below ideal values. Moreover, the raw responses exhibited statistical bias, rendering them unsuitable for direct use as cryptographic keys. Third, the performance evaluation was limited to Hamming distance analysis.

ioPUF+ represents a significant improvement over the original ioPUF. It measures both pull-up and pull-down resistors of the I/O pins to generate PUF responses, thereby extending the response length from 45 bits to 190 bits using the same number of pins. Additionally, ioPUF+ derives unique random identifiers for devices from biased PUF responses by employing BCH error correction and SHA-256 hashing techniques. The work also presents a comprehensive evaluation of PUF performance metrics, including reliability, stability, uniqueness, and bit-aliasing. Furthermore, ioPUF+ demonstrates the use of PUF in securing device-to-device communication through AES-based symmetric key cryptography.

This section reviews works related to {ioPUF+}, focusing on PUFs that exploit I/O characteristics and highlighting how they differ from our approach. Furthermore, we examine literature on PUF-based cryptographic key generation methods and compare them with the {ioPUF+} implementation.

\subsection{PUFs Related to I/Os}

The study by Basak et al. measures input currents corresponding to externally applied DC voltages at various I/O pins of an IC~\cite{basak_piraic}. Since the intrinsic resistances of pins vary across devices, the pins draw varying currents from the voltage source. The authors use these current measurements to create a unique device identifier for the IC. Our approach differs from that of Basak et al. in the following ways. 
First, unlike Basak et al., who focus on the measurement of intrinsic resistances of pins, we measure the resistances of the pull-up and pull-down resistors associated with the I/Os. Since the pull-up/down resistor values are significant compared to intrinsic pin resistances, the measurement apparatus required is less involved.
Second, in ioPUF+, the resistance measurements and generation of the PUF IDs are executed within the IC, eliminating the need for any external measurement apparatus. In contrast, the method in Basak et al. requires external voltage sources and high-precision ammeters connected to different pins of the SoC for input pin current or pin resistance measurements. Their method is less suited for PUF key generation in IoT nodes, which typically cannot accommodate costly, bulky measuring instruments. Therefore, Basak et al.'s method is more suited for lab-based device identification than real-world IoT node identification.

Vaidya et al. explores a PUF-based device identification scheme using I/O pins~\cite{vaidya2020gpio}. In their approach, I/O pins are driven to logic-low, and the difference in voltages between pairs of pins is measured. For the measurements, they connected I/O pins to a differential ADC in the IC through PCB traces external to the IC. Ideally, the ADC reading should be zero, considering both I/O pins are at a logic-low state. However, due to the parasitics of the I/O pin driver, the PCB traces, and the non-idealities of the ADC, the measurements will yield non-zero results that vary across the devices. Vaidya et al. utilized 32 pairs of I/O pins to formulate the PUF ID. Our work significantly differs from Vaidya et al. in several ways. First, our PUF  relies on the  pull-up/down resistors on the I/O pins, which are in place by design. In contrast, the PUF by Vaidya et al. relies on parasitics and non-idealities, elements not intentionally designed into the system. Reliance on parasitics for generating PUF keys presents several challenges. First, the parasitics are highly dependent on the device's operation. For example, the ground bounce effect significantly influences the logic low voltage on a device's I/O pins. When the device consumes significant current during its operation, the ground bounce effect kicks in, which can cause the I/O pin voltage to increase, thereby affecting the PUF keys. While the effects like ground bounce may also impact the measurement of pull-up/down resistors, the impact is substantially less due to their higher resistance values (several kiloohms) compared to the parasitic resistances (a few milliohms). Further, the characteristics of pull-up/down resistor values and their variations are thoroughly understood and documented by IC manufacturers, facilitating more reliable PUF design. Additionally, unlike our method, Vaidya et al. requires I/O pins to be dedicated solely to  PUF generation.

Vinko et al.~\cite{vinko_analogRC_PUF} present an analog RC PUF. The method uses the internal pull-ups of two I/O pins to charge two external capacitors, which are subsequently discharged through external resistors. The PUF response is obtained from measuring the discharge time of these two capacitors.
In comparison to our work, the design by Vinko et al.~has several shortcomings. Due to the limited number of I/O pins used, the analog RC PUF generates relatively short PUF responses, limiting its applicability in cryptographic applications such as secret-key generation. Vinko et al.~average multiple discharge-time measurements to obtain a reliable PUF response, which significantly increases the PUF response generation time. Furthermore, in their design, each I/O pin used for PUF measurements requires a dedicated external RC connection; therefore, the pin cannot be used for any other microcontroller function. In contrast, in our work, no passive components are connected to the pins whose pull-up/down resistors are measured. Finally, unlike our approach, the study by Vinko et al.~does not present detailed performance metrics or describe the process of generating PUF IDs from raw PUF responses.

\subsection{Secret Keys Using PUFs}

A key application of PUFs is their use in cryptographic operations. However, raw PUF responses cannot directly serve as cryptographic keys because they are noisy, vary with environmental conditions, and are not uniformly distributed, often exhibiting bias. To address these challenges, Dodis et al.~\cite{dodis2004fuzzy} proposed fuzzy extractors, which combine error correction to handle noisy PUF responses with hash functions to enhance randomness. Numerous studies have since built upon the work of Dodis et al. to implement PUF-based key generation systems. In this section, we review two representative works that adopt this approach and compare them with our implementation.

Suh et al.~\cite{puf_secretkey_gen1} present a PUF design for device authentication and cryptographic key generation. Their proposed ring oscillator PUF employs 1024 ring oscillators arranged in a 16×64 array on Xilinx Virtex-4 FPGAs. The design exploits manufacturing-induced frequency variations among ring oscillators to generate PUF responses. To improve the reliability of responses, the authors employ a 1-out-of-8 masking scheme that selects oscillator pairs with the maximum frequency separation for response generation. However, the implementation requires substantial hardware resources, as the 1024 oscillators consume significant area and power. This approach targets FPGA and ASIC platforms that offer abundant resources, in contrast to the low-cost, resource-constrained COTS microcontrollers considered in this work.

Maes et al.~\cite{pufky} present {PUFKY}, which is a PUF-based key generator that employs an optimized ring oscillator PUF with Lehmer-Gray encoding. PUFKY achieves high entropy density through a three-step process involving frequency normalization to remove bias, Lehmer encoding for stable ordering representation, and XOR-based entropy compression. This approach, however, increases implementation complexity, utilizing up to 1024 oscillators, concatenated error-correction codes, and a custom coprocessor for BCH decoding. PUFKY generates 128-bit keys with a resource usage of 17\% on an FPGA. In contrast, our approach utilizes only 6.4\% of resources on a low-cost, resource-constrained microcontroller.

In contrast to these FPGA/ASIC-centric methods, our implementation adopts the general fuzzy extractor concept~\cite{dodis2004fuzzy}, targeting COTS microcontrollers rather than specialized hardware. ioPUF+ leverages I/O pin pull-up and pull-down resistance variations to generate PUF responses. Lightweight BCH error correction combined with a hash function is applied to these responses to generate random yet reliable cryptographic keys. All these steps are performed in firmware and consume significantly less memory and power. Furthermore, we demonstrate the practical use of the generated cryptographic keys in enabling secure device-to-device communication.

%% file: myContents/section-4-system-design.tex
\section{System Design}
\label{sec_sysdesign}

{ioPUF+} comprises the following components: a PUF primitive based on the pull-up/down resistors of I/O pins, an algorithm for generating PUF responses using this primitive, a method for stabilizing the responses to derive unique device identifiers, and a hashing mechanism for generating secret keys from these identifiers. This section describes the design and operation of each of these components in detail.

\subsection{I/O-Based PUF Primitive}

\begin{figure*}[!htbp]
\centering
\includegraphics[width=0.6\linewidth]{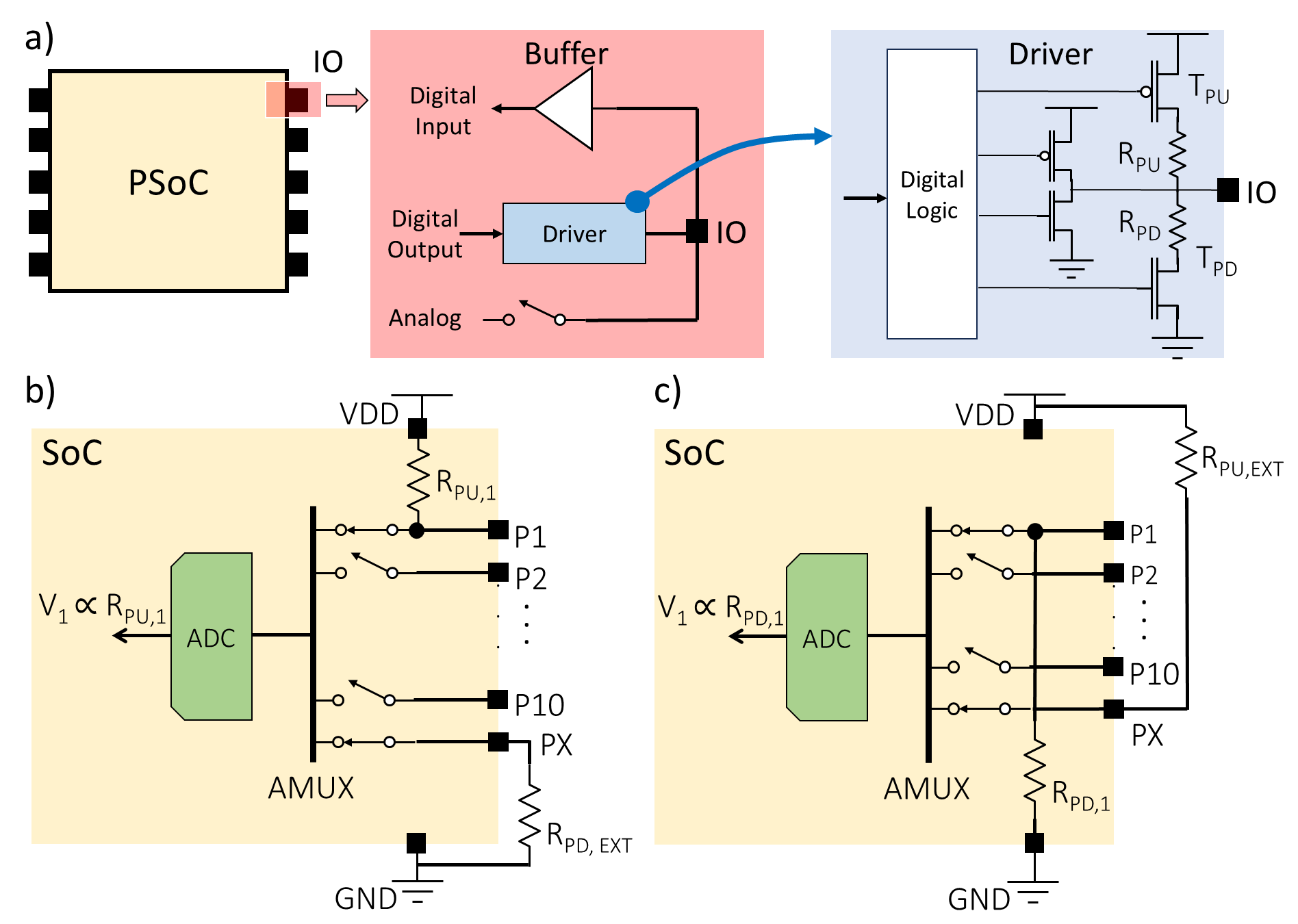}
\caption{a) I/O block of PSoC-5 from Infineon Technologies, highlighting the pull-up resistor $R_{PU}$ and pull-down resistor $R_{PD}$~\cite{infineon32bitPSoC}. b) Method of measuring pull-up resistor of $P1$ using an external pull-down resistor, $R_{PD, EXT}$, connected to $PX$. c) Proposed method of measuring pull-down resistors of $P1$ using an external pull-up resistor, $R_{PU, EXT}$.}
\label{fig_psoc_gpio_architecture}
\end{figure*}

Pull-up and pull-down resistors are present on the I/O pins of various ICs. Many of these ICs allow attaching/detaching of these resistors through firmware-based register writes. These resistors enable I/O pins for communication and logic operations. For instance, I2C-based communication requires pull-up resistor support for I/O pins~\cite{leens2009introduction}. Figure~\ref{fig_psoc_gpio_architecture}~(a) shows the I/O architecture of a microcontroller IC, PSoC-5, from  Infineon Technologies~\cite{infineon32bitPSoC}. In the figure, $R_{PU}$ and $R_{PD}$ represent the pull-up and pull-down resistors, respectively. These resistors can be attached/detached from the I/O by turning transistors $T_{PU}$ and $T_{PD}$ ON or OFF. PSoC-5 allows these transistors to be turned ON/OFF through firmware. The resistance values of these internal resistors vary between pins and across devices due to manufacturing process variations. {ioPUF+} utilizes these variations to produce unique PUF responses for the ICs.

{ioPUF+} generates PUF responses from the measurement of $R_{PU}$ and $R_{PD}$ for several I/O pins.
Figure~\ref{fig_psoc_gpio_architecture}(b) illustrates the proposed method for measuring $R_{PU}$ of an I/O pin $P_1$. A known reference resistor, $R_{PD,EXT}$, is connected to the I/O pin $P_X$. Subsequently, pins $P_1$ and $P_X$ are internally shorted, using AMUX\footnote{AMUX $\rightarrow$ Internal Analog Multiplexer in PSoC-5}, to form a resistor divider circuit. An internal ADC is then used to measure the output voltage of this divider. Equation~\ref{eqn_rpu_computation} shows the relationship between $R_{PU,1}$ and the measured ADC voltage $V_1$. Since $R_{PD,EXT}$ is known, the value of $V_1$ serves as an indicator of $R_{PU,1}$.
The same method can be used to measure the pull-up resistor of any other I/O pin. For instance, to measure $R_{PU,2}$ (associated with $P_2$), $P_1$ is disconnected from $P_X$, $P_2$ is shorted to $P_X$, and the ADC measures the corresponding divider output. The resulting ADC value $V_2$ is then used to estimate $R_{PU,2}$.

\begin{equation}
V_1 = \frac{V_{DD} \times R_{PD,EXT}}{R_{PU,1} + R_{PD,EXT}}
\label{eqn_rpu_computation}
\end{equation}

A similar approach is used to measure $R_{PD}$ for an I/O pin. Figure~\ref{fig_psoc_gpio_architecture}(c) illustrates the method for measuring $R_{PD}$ of pin $P_1$. A known reference resistor $R_{PU,EXT}$ is connected to pin $P_X$ in a pull-up configuration. Pins $P_1$ and $P_X$ are then shorted internally to form a resistor divider, and the ADC measures its output voltage. Since $R_{PU,EXT}$ is known, the measured voltage provides an estimate of $R_{PD,1}$.

{ioPUF+} collects twenty voltage readings corresponding to the pull-up and pull-down resistor measurements of ten I/O pins to generate PUF responses. In comparison, our previous work, {ioPUF}~\cite{ioPUF}, used only ten voltage readings corresponding to pull-up or pull-down resistor measurements, which limited the length and entropy of the responses.

\paragraph*{Note} For the measurement method described in this section to work, the ICs must support an IO architecture that enables $P_X$ to be connected or disconnected from every other IO. The architecture helps measure pull-up resistors associated with multiple IOs using a single dedicated IO, $P_X$, which connects to an external pull-down resistor. Many ICs equipped with ADCs that support multiple input channels already have the necessary switch fabric to enable these operations. However, it is feasible to implement ioPUF+ in ICs lacking this specific IO architecture. In such cases, every IO requiring the measurement of its pull-up resistor needs an external pull-down resistor connected to its terminal. A drawback of this method is the increased number of dedicated IOs, which require external resistors.

\paragraph*{Note} In order to facilitate pull-up and pull-down resistor measurements without any manual intervention, two pins, $PX_A$ and $PX_B$, were dedicated for connecting the external reference resistors, $R_{PU,EXT}$ and $R_{PD,EXT}$, respectively. Only one of these was connected to the AMUX at any given time.

\subsection{Generation of PUF Responses}

{ioPUF+} uses Algorithm~\ref{alg:ioPUF} to generate binary PUF responses from an IC. 

\begin{algorithm}[ht]
\caption{Method of Generating PUF Responses}
\label{alg:ioPUF}
\begin{algorithmic}[1]
\REQUIRE Voltages $V_1,V_2,\ldots,V_{N}$ from $N/2$ pull-up and $N/2$ pull-down resistors of pins $P_1,P_2,\ldots,P_{N/2}$.
\ENSURE A ${PUF\text{-}Response}$ of length $\binom{N}{2}$ bits.
\STATE Initialize an empty bit list ${B}$.
\FOR{$i = 1$ to $N-1$}
  \FOR{$j = i+1$ to $N$}
    \IF{$V_i < V_j$}
      \STATE Append $1$ to ${B}$.
    \ELSE
      \STATE Append $0$ to ${B}$.
    \ENDIF
  \ENDFOR
\ENDFOR
\STATE Concatenate all bits in ${B}$ to form ${PUF\text{-}Response}$.
\RETURN ${PUF\text{-}Response}$
\end{algorithmic}
\end{algorithm}

The algorithm processes the twenty voltage readings, $V_1, V_2, \ldots, V_{20}$, obtained from measuring the pull-up resistors ($V_1$ to $V_{10}$) and pull-down resistors ($V_{11}$$-$ $V_{20}$) of ten I/O pins $P_1, P_2, \ldots, P_{10}$. 

The PUF response is constructed by comparing every pair of these voltage readings and converting each comparison into a binary bit, which are then concatenated to form the final response. 
The twenty voltage readings generate a PUF response of length $\binom{20}{2} = 190$ bits, representing a four-fold increase in response length compared to our previous work, which produced a 45-bit PUF response~\cite{ioPUF}.
Since the voltage readings vary from one IC to another due to manufacturing variations, the PUF responses from two different ICs are also expected to differ. 

\paragraph*{Note}
A proof-of-concept SPICE simulation was performed to evaluate the effectiveness of using pull-up and pull-down resistors, along with Algorithm~\ref{alg:ioPUF}, for generating PUF responses. The simulations were conducted in LTspice~\cite{ltspice}, modeling the internal resistors of the I/O pins, their inter- and intra-device variations, and their temperature and voltage dependencies. The  study provided insights into the uniqueness of the PUF responses and demonstrated their reliance on model parameters, particularly on ADC resolutions. The simulation setup, results, and corresponding observations are described in detail in our earlier work, {ioPUF}~\cite{ioPUF}.

\subsection{Derivation of PUF ID and Secret Key}
\label{sec_sys_design_secretkeys}

Secret keys are required for many cryptographic applications, such as data encryption, message authentication, and secure device-to-device communication. For use in cryptographic applications, secret keys must be both stable and random.
A secret key is considered stable if the key generated by a device remains consistent across diverse operating conditions, such as temperature and supply voltage fluctuations, and over time. Stable secret keys are essential in cryptographic applications, as a key that changes unpredictably can lead to decryption failures or data packet losses in communication links. 
A secret key is considered random if the key generated by a device is unpredictable. Such keys are less prone to statistical and brute-force attacks.

\begin{figure}[!htbp]
\centering
\includegraphics[width=1\linewidth]{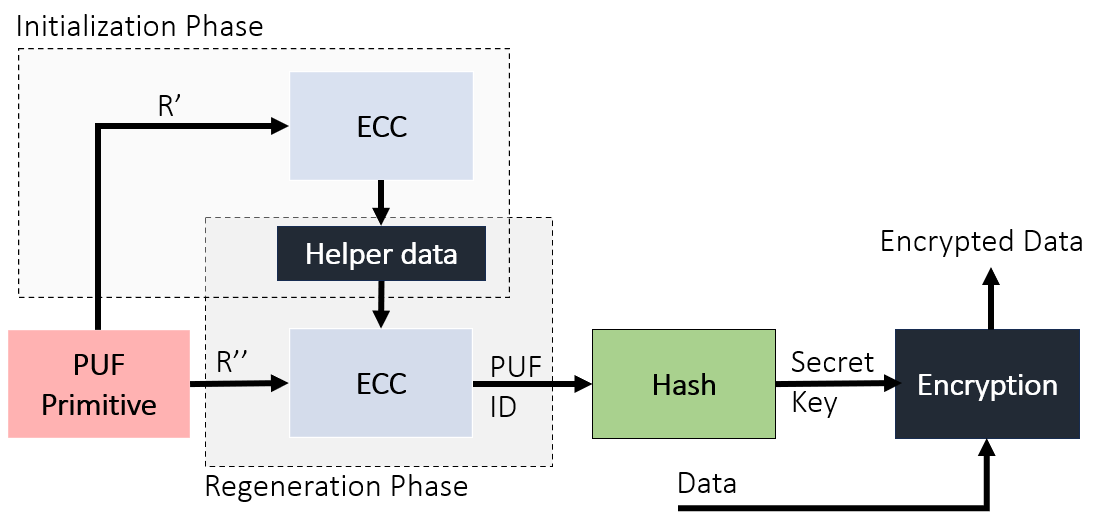}
\caption{Derivation of the PUF ID and secret key from PUF responses using BCH error-correcting codes (ECC) and SHA-256 hashing, along with an example application of the derived secret key for data encryption.}
\label{fig_sys_design_id_key_gen}
\end{figure}

The following sections describe the method used by {ioPUF+} to generate a secret key within a device. The first part explains the derivation of a unique device identifier from PUF responses using BCH error-correcting codes (ECC). The second part details the use of the SHA-256 hash function to randomize this identifier and generate a secret key~\cite{fuzzy_extractor,puf_secretkey_gen2}. Figure~\ref{fig_sys_design_id_key_gen} presents a schematic of these operations.

PUF responses from a device vary with environmental noise, temperature, and supply voltage~\cite{porlapothula2025cappuf} (see Section~\ref{sec_evaluation}). Hence, a raw PUF response cannot be used directly as a unique device identifier; it must first be stabilized. Error-correcting codes (ECC) are employed for this purpose. Among the various ECC schemes reported in the literature, {ioPUF+} uses the Bose–Chaudhuri–Hocquenghem (BCH) code. BCH codes are chosen because they are lightweight and suitable for systems such as {ioPUF+}, where the bit error rate is relatively low~\cite{bch_ecc_devadas}.

Figure~\ref{fig_sys_design_id_key_gen} presents the two operational phases of BCH: the initialization phase and the regeneration phase.

The initialization phase is executed once per device, typically during enrollment. In this phase, a PUF response, R', from the device is recorded. This response is enrolled as the unique identifier of the device, the PUF ID. The ID is processed by the BCH encoder to generate helper data, which is then stored in the non-volatile memory of the device, such as EEPROM~\cite{cypressEEPROMv2_10}. 

The regeneration phase is executed whenever the PUF ID is required. In this phase, a fresh PUF response, R'', is recorded. This response may differ from the enrolled ID by a small number of bits due to noise or environmental variations. Using the stored helper data, the BCH decoder corrects these errors and reconstructs the original PUF ID.

However, the resultant PUF ID may not be truly random and therefore cannot be used directly as a secret key. To address this, {ioPUF+} randomizes the PUF ID using the SHA-256 hash function~\cite{sha256} to generate a secret key. The secret key may then be subsequently used for cryptographic applications, such as data encryption, as shown in Figure~\ref{fig_sys_design_id_key_gen}. The encrypted data can then be transmitted or stored securely.

\paragraph*{Note} Although the helper data is stored in non-volatile memory and may be accessible to an attacker, it alone does not reveal information about the PUF responses or the PUF ID. The helper data contains only the information necessary for the BCH decoder to correct errors in PUF responses. Even if an adversary obtains the helper data, they cannot reconstruct the original PUF ID without access to the PUF responses, which are generated only at runtime and are never stored in non-volatile memory. Furthermore, the helper data is a many-to-one function of the PUF response, meaning that multiple distinct responses can yield the same helper data~\cite{puf_secretkey_gen2}. Therefore, reverse-engineering PUF responses or the PUF ID from the helper data of a device is computationally difficult.

%% file: myContents/section-5-results.tex
\section{Evaluation of PUF Metrics}
\label{sec_evaluation}

In this section, we evaluate the PUF responses using standard PUF performance metrics, namely reliability, stability, uniqueness, uniformity, and bit-aliasing. We also present the results of temperature and supply-voltage variation tests to assess the robustness of the PUF responses.

The first part of this section describes the experimental setup, including the implementation details of ioPUF+ on the PSoC-5 microcontroller. The second part outlines each PUF metric and its evaluation methodology. Finally, we present and discuss the experimental results.

\subsection{Experimental Setup}
\label{sec_exp}
\label{sec_exp_setup}

To evaluate ioPUF+, we first implemented the PUF on the PSoC-5 microcontroller IC~\cite{infineon32bitPSoC} from Infineon Technologies. We then developed a dedicated experimental setup around the PSoC-5 to characterize the performance of the PUF.

\begin{figure*}[!htbp]
\centering    
\includegraphics[width=0.9\linewidth]{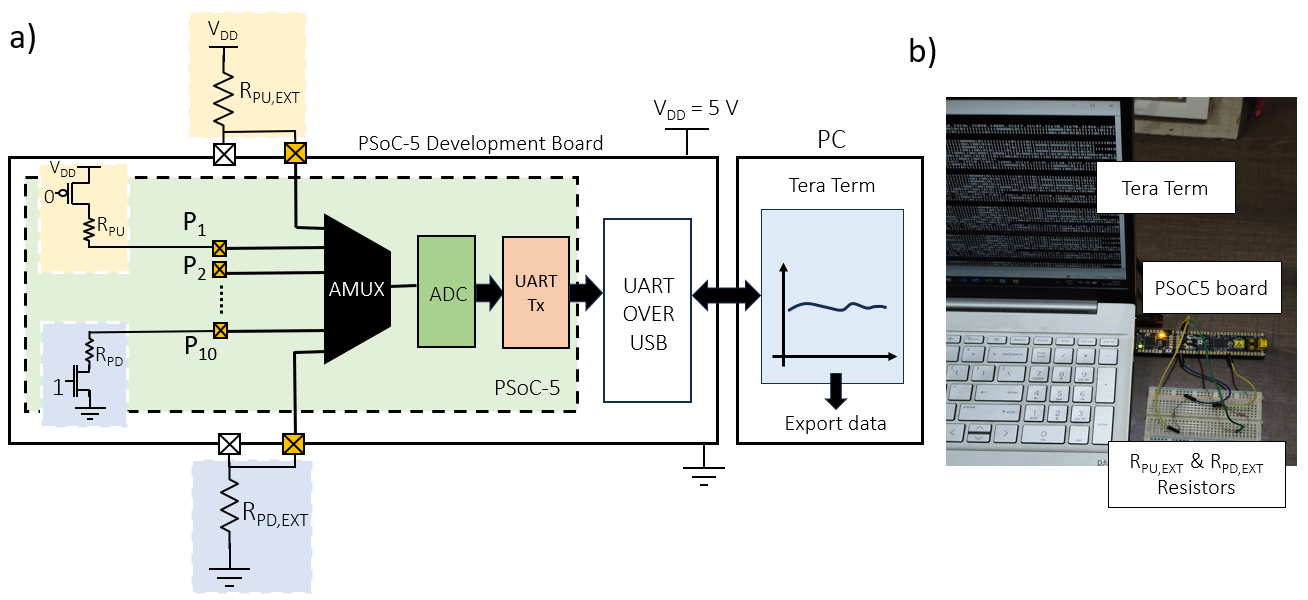}
\caption{a) Schematic of the experimental setup consisting of the PSoC development board, CY8CKIT-059, connected to a PC via a USB port. The ADC measures the voltage across the external resistors $R_{PU,EXT}$ and $R_{PD,EXT}$ for different AMUX select settings, allowing measurement of both pull-up and pull-down resistor values of the GPIOs: $P_1$ to $P_{10}$. Measurements are transferred to the PC in UART format and received using the Tera Term serial terminal. Subsequently, the readings are processed to generate the PUF responses. b) Picture of the experimental setup highlighting the Tera Term serial terminal, PSoC development board, and the external resistor connection.}
\label{fig_setup}
\end{figure*}

Figure~\ref{fig_setup}~(a) shows the schematic of the experimental setup. The setup consists of a PSoC-5 board, CY8CKIT-059~\cite{infineonCY8CKIT059Infineon}, connected to a PC via a USB port. The CY8CKIT-059 board includes two ICs: the PSoC-5, which serves as the device under test, and a UART-over-USB interface used to transfer data from the PSoC-5 to the PC. We used this experimental setup to implement and evaluate four ioPUF+ configurations: (1) ioPUF+ using only the pull-up resistors, (2) ioPUF+ using only the pull-down resistors, (3) ioPUF+ using both pull-up and pull-down resistors, and (4) ioPUF+ using both pull-up and pull-down resistors with SHA-256 hashing.

For configuration (1), we implemented the pull-up resistor measurement circuit shown in Figure~\ref{fig_psoc_gpio_architecture}~(b) in the PSoC-5. The implementation used ten I/O pins in resistive pull-up mode ($R_{PU}$ of $5.6\text{K}\Omega$), an analog multiplexer (AMUX), an ADC, and an external resistor, $R_{PD,EXT}$, of $5.6\text{K}\Omega$. We used the breadboard shown in Figure~\ref{fig_setup}~(a) to connect the external pull-down resistor $R_{PD,EXT}$ and to wire the output of the MUX to the input of the ADC. We connected the I/O pins sequentially to $R_{PD,EXT}$ by configuring the selection bit of the multiplexer in the PSoC-5 firmware. Connecting I/O $P_i$ to $R_{PD,EXT}$ resulted in a resistive-divider network, yielding a voltage across $R_{PD,EXT}$, given by:
\[
V_i = \frac{VDD \times R_{PD, EXT}}{R_{PU,i} + R_{PD, EXT}}.
\]

We used the ADC to measure the voltage $V_i$. For the measurements, the input range of the ADC was set to 0–5V, and the resolution was set to 16 bits, yielding a voltage resolution of approximately $76\,\mu\text{V}$. Since we used ten I/O pins, the arrangement facilitated the measurement of ten different voltages, $V_1$ to $V_{10}$, which are representative of the values of the pull-up resistors corresponding to the ten I/Os. We programmed the PSoC-5 to continuously measure $V_1$ to $V_{10}$ and stream the data via UART at a baud rate of 115200bps to the PC through the UART-Over-USB IC. On the PC side, we used the Tera Term serial terminal to receive and record the data via the PC's COM port. The PUF responses for the PSoC-5 device were generated from these voltage values on the PC side.

Configuration (2) is similar to configuration (1), with the following differences. For configuration (2), we implemented the pull-down resistor measurement circuit shown in Figure~\ref{fig_psoc_gpio_architecture}~(c) in the PSoC-5. We used the breadboard shown in the figure to connect the external pull-up resistor $R_{PU,EXT}$ and to wire the output of the MUX to the input of the ADC. In this configuration, voltage values $V_1$ to $V_{10}$ represented the values of the pull-down resistors corresponding to the ten I/O pins of the PSoC-5. These voltage readings were streamed to the PC. The PUF responses for the PSoC-5 device were generated from these voltage values on the PC side.

Configuration (3), as shown in Figure~\ref{fig_setup}, is implemented by combining methods (1) and (2) in the PSoC-5. In this configuration, first, the pull-up resistors of the 10 I/O pins of the PSoC-5 device were measured using the circuit configuration shown in Figure~\ref{fig_psoc_gpio_architecture}~(b). The pull-up resistor values were recorded as voltage readings $V_1$ to $V_{10}$. Next, the pull-down resistors of the 10 IOs of the PSoC-5 were measured using the circuit configuration shown in Figure~\ref{fig_psoc_gpio_architecture}~(c). The pull-down resistor values were recorded as voltage readings $V_{11}$ to $V_{20}$. These voltage readings, $V_1$ to $V_{20}$, were then transmitted to the PC. The PUF responses of the PSoC-5 device were generated from these 20 voltage readings on the PC. This configuration achieved PUF responses of length $\binom{20}{2} = 190$ bits, higher than the PUF responses' length of $\binom{10}{2} = 45$ bits achieved in configurations (1) and (2).

Configuration (4) is similar to (3), with the following addition: a hash function is applied to the PUF responses to improve their randomness.

In all configurations, we recorded voltage measurements from the PSoC-5 device over a 30-minute period. Each set of voltage readings was used to generate a PUF response, resulting in a series of PUF responses for the device over the $\approx$30 minutes. These measurements were used to quantify the drift in PUF responses for the device over time. Furthermore, we repeated the experiment on 32 different CY8CKIT-059 boards, i.e., 32 different PSoC-5 devices. The experiments were performed at an ambient temperature of $\approx~29\,^{\circ}\mathrm{C}$. Figure~\ref{fig_setup}~(b) shows the experimental setup with the CY8CKIT-059 connected to the PC. 

\subsection{Standard PUF Metrics}
\label{sec_PUF_Metrics_Definition}

In the following subsections, we describe the standard PUF metrics~\cite{maiti2013systematic} and the corresponding evaluation methodology used in this work.

\subsubsection{Reliability}

Reliability refers to a PUF’s ability to consistently reproduce the same response across multiple measurements from the same device over time, under similar operating conditions. It is evaluated using the mean intra-device Hamming distance, denoted as $\overline{HD}_{\text{intra}}$:

\begin{equation}
\overline{HD}_{\text{intra}}(\%) = \frac{1}{N} \sum_{i=1}^{N} 100 \times \frac{d_H(P_1, P_i)}{L}
\label{eqn_mean_intra_HD}
\end{equation}

\begin{equation}
\text{Reliability}~(\%) = 100\% - \overline{HD}_{\text{intra}}
\label{eqn_reliability}
\end{equation}

where:

\begin{itemize}
    \item $d_H(P_1, P_i)$ represents the Hamming distance between the reference PUF response $P_1$ and the $i$-th PUF response $P_i$, which is measured from the same device at a different time.
    \item $N$ represents the number of PUF responses obtained from the device through repeated measurements.
    \item $L$ denotes the number of bits in each PUF response.
\end{itemize}

\subsubsection{Stability}
Stability measures a PUF's ability to reproduce consistent PUF responses when subjected to varying operating conditions such as changes in temperature or supply voltage. 

Stability is evaluated as follows. First, a reference PUF response is recorded from a device under typical operating conditions. The same device is then tested under a range of temperature and voltage settings, to record new PUF responses. The stability is computed by measuring the percentage of bit flips between the reference and the subsequent PUF responses. 

\subsubsection{Uniqueness}

Ideally, the PUF responses from different devices should be as dissimilar as possible to  distinctly identify them. The uniqueness metric measures the ability of a PUF to differentiate devices based on the generated PUF response. 

Uniqueness is computed by averaging the Hamming distances between all possible pairs of PUF responses from different devices, i.e., the mean inter-device Hamming distance, denoted as $\overline{HD}_{\text{inter}}$:

\begin{equation}
\overline{HD}_{\text{inter}}(\%) = \frac{1}{\binom{K}{2}} \sum_{i=1}^{K-1} \sum_{j=i+1}^{K} 100 \times \frac{d_H(P_i, P_j)}{L}
\label{eqn_mean_inter_HD}
\end{equation}

\noindent where:
\begin{itemize}
    \item $d_H(P_i, P_j)$ is the Hamming distance between the PUF responses generated by devices $P_i$ and $P_j$.
    \item $K$ represents the number of devices in the experiment.
    \item $L$ is the length of each PUF response in bits.
    \item $\binom{K}{2} = \frac{K(K-1)}{2}$ is the number of unique device pairs formed from $K$ devices.
\end{itemize}

\subsubsection{Uniformity}

Uniformity evaluates how evenly 0s and 1s are distributed in a PUF response, which reflects the bit-level randomness of the PUF response. A uniform distribution ensures that the PUF does not favor either binary value, making the PUF response statistically unbiased and difficult to predict.

Uniformity is calculated by determining the fraction of bits set to '1' in the PUF response. The result is expressed as a percentage, $U_i(\%)$, representing the percentage of 1s in the PUF response for device $i$:

\begin{equation}
U_i(\%) = \frac{1}{L} \sum_{j=1}^{L} p_{i,j} \times 100
\label{eqn_uniformity}
\end{equation}

\noindent where:
\begin{itemize}
    \item $p_{i,j}$ is the $j$-th bit of the PUF ID generated by device $i$.
    \item $L$ is the number of bits in each PUF ID
    
\end{itemize}

\subsubsection{Bit-Aliasing}

Bit-aliasing measures how consistently a specific bit position appears as either 0 or 1 across PUF responses from different devices. It helps assess whether any particular bit position in the PUF output is biased across a population of devices. Ideally, each bit position should have an equal likelihood of being 0 or 1.

This metric is calculated by averaging the value of a given bit position across all devices. The result is expressed as a percentage, $B_j(\%)$, indicating the proportion of devices where the $j$-th bit is 1. The ideal value of the metric is 50\%, indicating no bias, i.e., each bit position is equally likely to be 0 or 1 across devices.

\begin{equation}
B_j(\%) = \frac{1}{K} \sum_{i=1}^{K} P_{i,j} \times 100
\label{eqn_bitaliasing}
\end{equation}

\noindent where:
\begin{itemize}
    \item $P_{i,j}$ is the $j$-th bit of the PUF response for device $i$.
    \item $K$ is the total number of devices in the evaluation.
\end{itemize}

\subsection{Results}

\subsubsection{Reliability, Uniqueness, and Randomness}

\begin{figure*}[!htbp]
\centering    
\includegraphics[width=1\linewidth]{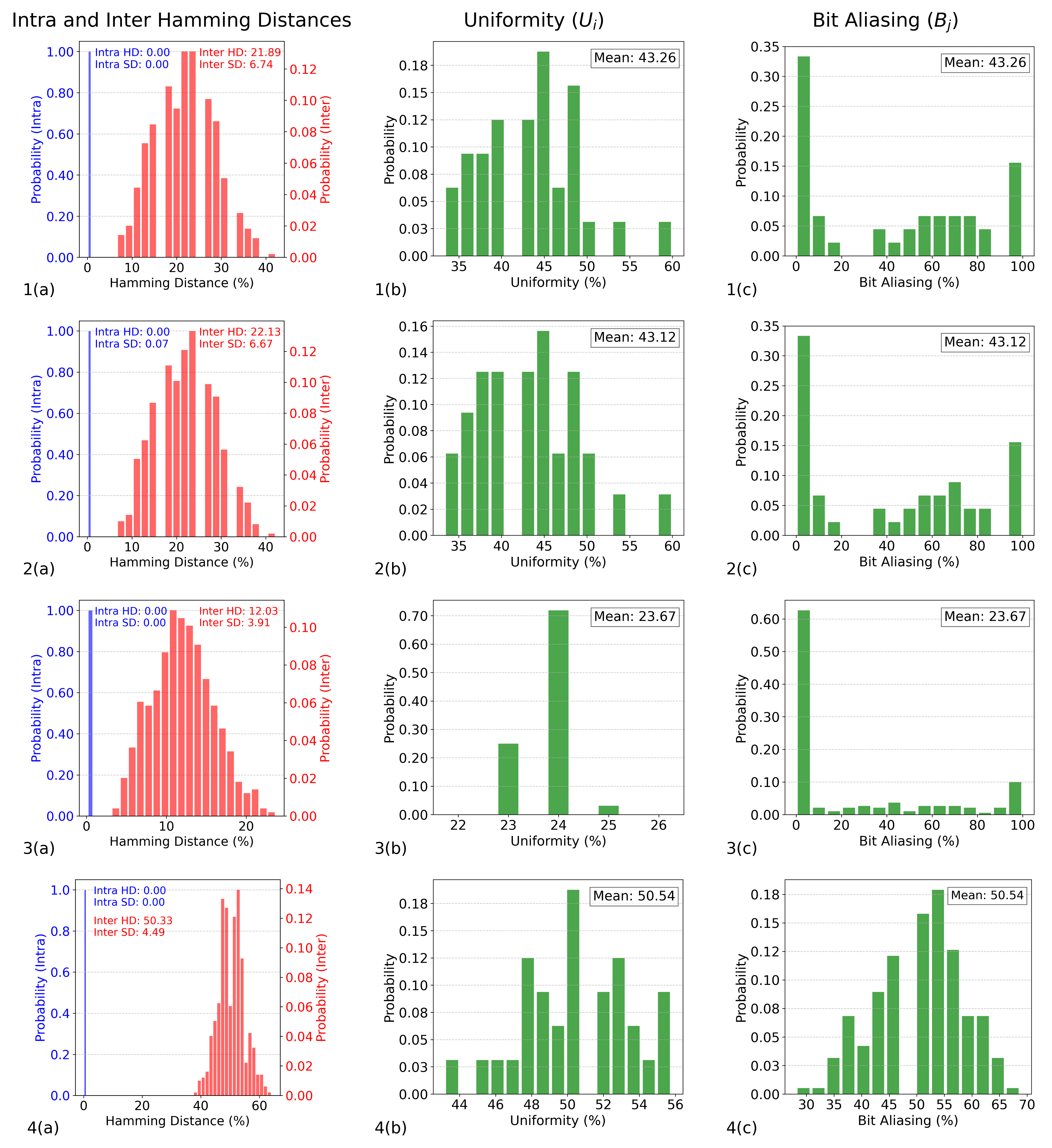}
\caption{Plot of standard PUF metrics for the four ioPUF+ configurations: (1) using only the pull-up resistors, (2) using
only the pull-down resistors, (3) using both pull-up
and pull-down resistors, and (4) using both pull-up
and pull-down resistors with SHA-256 hashing.}\label{fig_metrics_plots}
\end{figure*}

Figure~\ref{fig_metrics_plots} presents the PUF metrics evaluated for the four ioPUF+ configurations described in Section~\ref{sec_PUF_Metrics_Definition}. Specifically, the figure includes histogram plots of the intra- and inter-Hamming distances, uniformity, and bit aliasing for the four ioPUF configurations. Table~\ref{tbl_metric_values} presents the computed metric values for reliability, uniqueness, uniformity, and bit aliasing. In Table~\ref{tbl_metric_values}, the average uniformity and bit-aliasing values are identical, as they compute the same overall fraction of '1' bits when averaged 
across all devices and bit positions, however, their per-device and per-bit 
distributions differ significantly (see Figure~\ref{fig_metrics_plots}).

For Configurations (1), ioPUF+ using only pull-up resistors, and (2), ioPUF+ using only pull-down resistors, the intra-Hamming distances are observed to be 0, indicating that the PUF response from a device remains stable over time, as expected of an ideal PUF. The inter-Hamming distances, which quantify the uniqueness of the PUF responses across different devices, are 21.89\% and 22.13\% for the pull-up and pull-down configurations, respectively. These values are significantly lower than the 50\% expected of an ideal PUF.

The mean uniformity measured for both configurations exceeds 40\% but remains below the ideal value of 50\%, indicating an imbalance in the distribution of ‘0’s and ‘1’s in the PUF responses. {The bit-aliasing analysis reveals significant deviations from the ideal behavior, with distributions heavily skewed toward 0\% and 100\% rather than clustering around the ideal value of 50\%. This clustering at the extremes indicates that specific bit positions consistently produce identical values (either all 0s or all 1s) across different devices, which significantly reduces the randomness and entropy of the PUF responses. For an optimal PUF, the bit-aliasing distribution should be concentrated around 50\%, indicating that each bit position has an equal probability of being 0 or 1 across the device population.}

A limitation of Configurations (1) and (2), ioPUF+ using either pull-up resistors or pull-down resistors, is their restricted PUF response length. Using 10 GPIOs results in a PUF response of length $\binom{10}{2} = 45$ bits, which falls short of the 128-bit length required by many modern cryptographic algorithms. Although extending the PUF response length by using more GPIOs is possible, it is often impractical due to the limited number of I/O pins available on cost-sensitive microcontrollers typically used in wearable and IoT devices. To overcome this limitation, Configuration (3) employs both pull-up and pull-down resistors, enabling the generation of $\binom{20}{2} = 190$ bits PUF response using the same 10 GPIOs.

{
For Configuration (3) (using both pull-up and pull-down resistors), the intra-Hamming distance remains 0, as expected of an ideal PUF. However, the mean inter-Hamming distance, uniformity, and bit-aliasing deviate from their ideal values, measuring 12.03\%, 23.6\%, and 23.6\%, respectively. These metrics are lower compared to Configurations (1) and (2), as presented in Table~\ref{tbl_metric_values}. This behavior suggests that combining both resistor types introduces systematic biases in the PUF response generation. Despite these deviations, the non-overlapping distributions of inter and intra-Hamming distances confirm that the PUF responses remain different across the devices, as shown in figure \ref{fig_metrics_plots}-3(a).}

To mitigate the biasing observed in Configuration (3), Configuration (4) applies a SHA-256 hash function to the PUF response generated in Configuration (3) \cite{sec_gen_biased_puf}. The hashed output is truncated to 190 bits to match the PUF response length of Configurations (1), (2), and (3), for a fair comparison of their metrics. The PUF metrics for Configuration (4) show significant improvement: the inter-Hamming distance increases to 50.29\%, the mean uniformity approaches the ideal 50\%, and the bit-aliasing plot forms a near-perfect parabolic curve centered at 50.54\%. These results indicate that applying hashing effectively mitigates the systematic biases introduced when combining both resistor types, resulting in a PUF response with substantially improved randomness and suitability for cryptographic applications.

\begin{table*}[!htbp]
\centering
\caption{Evaluated PUF metric values for four different ioPUF+ Configurations}
\label{tbl_metric_values}
\begin{tabular}{|p{7cm}|c|c|c|c|}
\hline
\textbf{ioPUF+ Configuration} & \textbf{Reliability (\%)} & \textbf{Uniqueness (\%)} & \textbf{Uniformity (\%)} & \textbf{Bit-Aliasing (\%)} \\
\hline
Using only the pull-up resistors & 100.00 & 21.89 & 43.26 & 43.26 \\
\hline
Using only the pull-down resistors & 100.00 & 22.13 & 43.12 & 43.12 \\
\hline
Using both pull-up and pull-down resistors & 100.00 & 12.03 & 23.67 & 23.67 \\
\hline
Using both pull-up and pull-down resistors with hashing & 100.00 & 50.29 & 50.54 & 50.54 \\
\hline
\end{tabular}
\end{table*}

\subsubsection{Stability}
\label{sec_stability_experiments}

Stability of PUF responses was evaluated across varying temperatures and supply voltages, as described in~\cite{lee2020rc}. For the experiments, Configuration (3), the ioPUF+ implementation using both pull-up and pull-down resistors, was used.

\paragraph{Impact of Temperature}

\begin{figure*}[!htbp]
\centering
\includegraphics[width=0.95\linewidth]{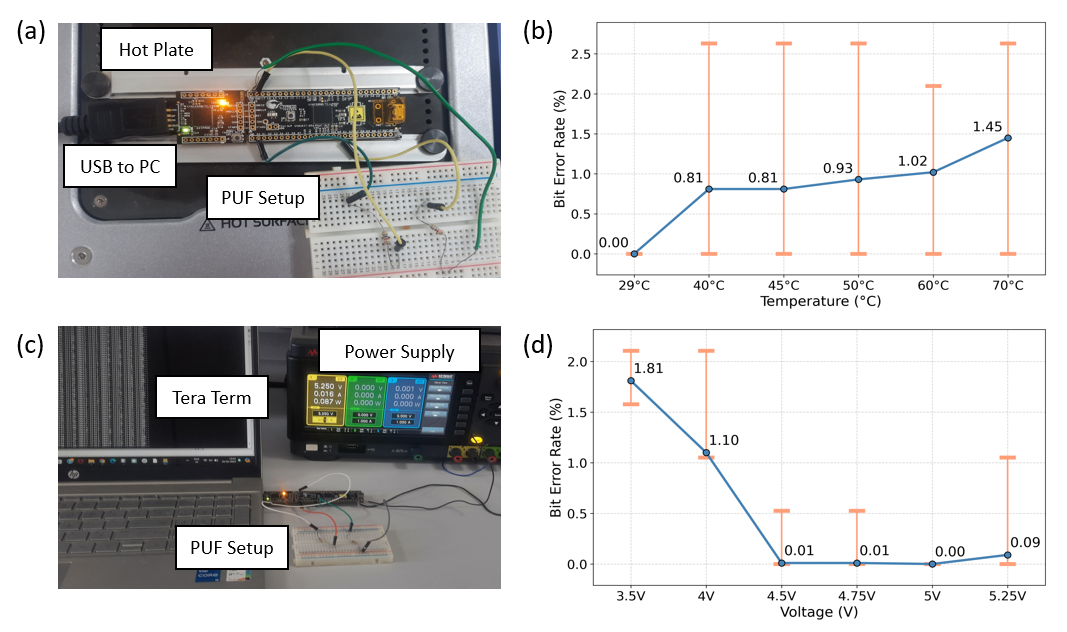}
\caption{Experimental setup and results for the stability tests of PUF responses. 
(a) Experimental setup for temperature variation using a hotplate. 
(b) Bit Error Rate (BER) of PUF responses vs.\ temperature. 
(c) Experimental setup for supply-voltage variation. 
(d) Bit Error Rate (BER) of PUF responses vs.\ supply voltage.}
\label{temp_volt_plots}
\end{figure*}

Figure~\ref{temp_volt_plots}~(a) shows the experimental setup used to evaluate stability across varying temperatures. We implemented the PUF on the PSoC-5 device of a CY8CKIT-059 board. The board was placed on a hot plate to vary the temperature of the PSoC-5. The die temperature was measured using the PSoC-5's built-in on-chip temperature sensor, which was configured in firmware to report the internal die temperature. Firmware was developed on the PSoC-5 to measure voltages $V_1$ to $V_{20}$, corresponding to the pull-up and pull-down resistor values, at regular intervals and transmit the data along with the die temperature of PSoC-5 to the connected PC via UART over USB.

Initially, a reference PUF response was generated by averaging 10,000 samples of $V_1$ to $V_{20}$ at room temperature (29°C). The temperature was then increased in gradual steps using the hot plate, and PUF responses were recorded at 40°C, 45°C, 50°C, 60°C, and 70°C. Stability was determined by measuring the number of bit flips in the PUF responses at each temperature relative to the reference PUF response.

Figure~\ref{temp_volt_plots}~(b) presents the stability of the PUF responses over varying temperatures. The vertical bars represent the minimum and maximum stability values observed across multiple measurements at each temperature point. The results indicate that the PUF responses remain relatively stable across a wide temperature range. The maximum mean stability remained below 1.5\%, and the worst-case stability observed was 2.63\%, corresponding to five bit flips.

\paragraph{Impact of Supply Voltage}

Figure~\ref{temp_volt_plots}(c) shows the experimental setup used to evaluate the stability of PUF responses under varying supply voltages. The setup consisted of a PSoC-5 device implementing the PUF, which continuously transmitted voltages $V_1$ to $V_{20}$ at regular intervals to a connected PC.

For the experiments, the USB power supply to the PSoC-5 was bypassed and replaced with an external laboratory-grade voltage source. The voltage source was configured to deliver supply voltages of 3.50V, 4.00V, 4.50V, 4.75V, 5.00V, and 5.25V in discrete steps. Initially, a reference PUF response was generated by averaging 10,000 samples of $V_1$ to $V_{20}$ at a supply voltage of 5.00V. The supply voltage was then adjusted in gradual steps, and the corresponding PUF responses were recorded at each voltage setting. Stability was determined by measuring the number of bit flips in the PUF responses at each voltage level relative to the reference PUF response.

Figure~\ref{temp_volt_plots}~(d) presents the stability of the PUF responses across the tested voltage range. The vertical bars represent the minimum and maximum stability values observed across multiple measurements at each voltage setting. The worst-case stability was found to be only 2.10\%, corresponding to approximately four bit flips, indicating that the PUF responses remain relatively stable across varying supply voltages.

%% file: myContents/section-5b-applications.tex
\section{Secret Key Generation and its Application}
\label{sec_applications}
In this section, we first describe the implementation details for deriving the secret key from the PUF responses on the PSoC-5, following the method outlined in Section~\ref{sec_sys_design_secretkeys}. The subsequent part demonstrates the use of the derived secret key to encrypt data transferred from the PSoC-5 to a connected PC. Finally, we evaluate the resource and energy consumption associated with secret-key generation and its use in encrypting data transfers.

\subsection{Secret Key Generation}

For generating a secret key from PUF responses, we followed the method described in Section~\ref{sec_sys_design_secretkeys}. First, the PUF responses were processed using a BCH error-correcting code to generate a unique PUF ID for the PSoC-5. Next, the secret key was derived by applying the SHA-256 hash  to the PUF ID.

\subsubsection{BCH ECC for PUF ID Generation}

The BCH implementation on the PSoC-5 was realized by porting the open-source BCH library from~\cite{parrot_bch}.

\paragraph{BCH Parameter Selection}

A BCH code is defined by three parameters \((n, k, t)\) and is represented as BCH\((n,k,t)\), where \(n\) represents the codeword length, \(k\) is the message length, and \(t\) defines the maximum number of correctable errors.  

The error correction capability \(t\) was set to 5. This choice was based on voltage and temperature variation tests on the PUF responses, which showed a worst-case error of about five bits (see Section~\ref{sec_stability_experiments}).

Next, the minimum values of \(n\) and \(k\) were selected to satisfy the following constraints~\cite{dodis2004fuzzy}: 
\begin{enumerate}
\item $n \geq k + \log_{2}(n+1) \cdot t$
\item $k \geq d$, where \(d\) is the length of the PUF response, 190 bits in our case.  
\end{enumerate}

These requirements were satisfied with \(n = 255\) and \(k = 215\), resulting in the BCH configuration of \(\text{BCH}(255,215,5)\). We selected these parameters because they provide sufficient error correction (t=5 bits) for our observed bit error rates while maintaining low computational overhead on the resource-constrained PSoC-5.

\paragraph{BCH Implementation}

The BCH implementation was carried out in two phases: Initialization and Regeneration. During the initialization phase, executed only once per device lifetime, the 190-bit PUF response, recorded as the PUF ID, was passed through the $\text{BCH}(255,215,5)$ encoder. The encoder generated helper data $H$ of size $(n-k) = 40$ bits, which was stored in the EEPROM of the PSoC-5. This helper data was then used by the $\text{BCH}(255,215,5)$ decoder in the subsequent regeneration phase to generate the unique device identifier, the PUF ID, from erroneous PUF responses.

\begin{figure*}[!htbp]
\centering    
\includegraphics[width=1\linewidth]{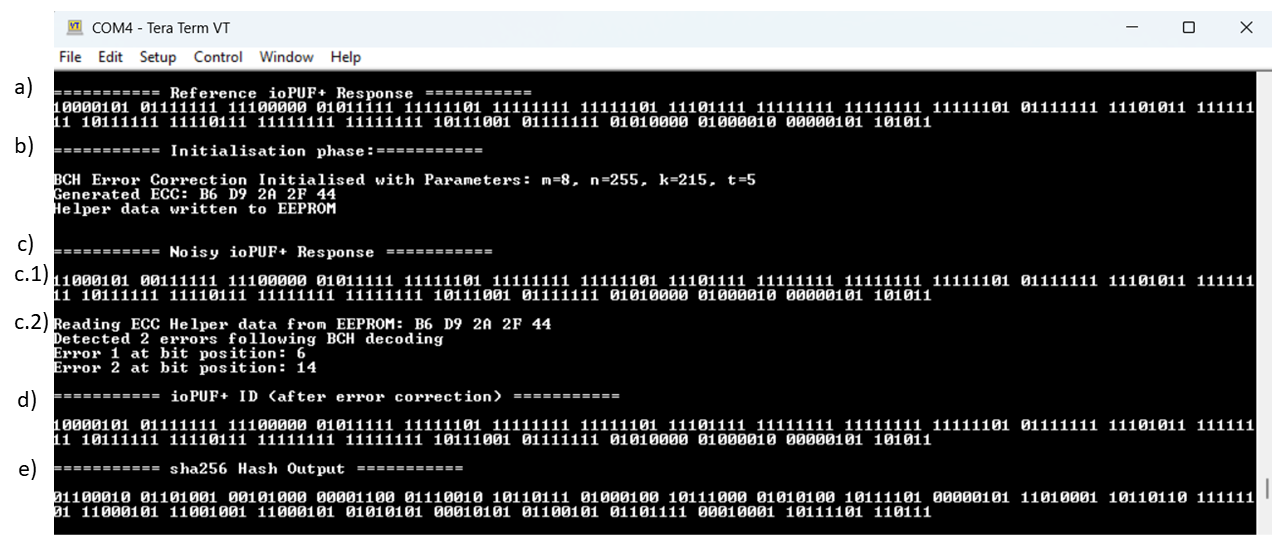}
\caption{Terminal output demonstrating BCH error correction and SHA-256 hashing on the PSoC-5.
(a) PUF response generated by the device during the BCH initialization phase, recorded as the PUF ID.
(b) BCH error-correction parameters and generated helper data after BCH encoding.
(c) Regeneration phase:
(c.1) Noisy PUF response,
(c.2) Detection and correction of bit errors using stored helper data.
(d) Corrected PUF response, PUF ID, after BCH decoding.
(e) SHA-256 hash output derived from the PUF ID, representing the secret key.}
\label{bcc_implementation}
\end{figure*}

 BCH implementation was evaluated on PSoC-5 by manually injecting bit errors into the PUF responses. We noted that the BCH decoder successfully identified and corrected these errors using the helper data  from the EEPROM. Figure~\ref{bcc_implementation} presents the results of this evaluation.

\subsubsection{SHA-256 Hashing for Secret Key Generation}

The PUF ID generated from the PUF response using BCH was further processed with a SHA-256 hash function to generate the secret key. The hash function implementation on the PSoC-5 utilized Intel's TinyCrypt SHA-256 library~\cite{sha256}, which was ported to the ARM Cortex-M3 architecture of the PSoC-5. 

The hash function accepted the PUF ID as input and generated a 256-bit output that served as the secret key. Hashing decorrelated the PUF response from the secret key, preventing statistical analysis attacks. The 256-bit hash output could be used directly or truncated to 128/192 bits depending on the encryption method employed (e.g., AES-128/192/256). 

Figure~\ref{bcc_implementation} presents the hash output in both hexadecimal and binary, as printed on the terminal by the PSoC-5.

\subsection{Encrypting Device-to-Device Communication}

Encrypting data transfers using secret keys is a common method to ensure confidentiality. Using PUF-derived secret keys for encryption offers a significant advantage: since the keys are generated at runtime from intrinsic and unclonable hardware features of the devices, they are difficult to extract through physical attacks.  

In this section, we showcase the use of the PUF-derived secret key to secure device-to-device communication. Specifically, we show the encrypted transfer of ECG data stored in the PSoC-5 to a PC. The PSoC-5 and the PC are physically connected and exchange data via UART over USB. The PSoC-5 stores a few samples of ECG data, which are first encrypted using the secret key and then transferred to the PC. The AES-128 encryption algorithm is used for this demonstration~\cite{tinyAESc}.  The ECG signal data is first divided into 16-byte blocks and padded using PKCS\#7 when the last block is smaller than 16 bytes. Each block is encrypted using AES-128 in ECB mode. The first  128 bits of the 256-bit secret key are used for AES-128 encryption. The encrypted blocks are transmitted sequentially over the UART to the PC via the USB interface. 

At the PC side, a Python script receives each 16-byte encrypted block through the COM port. The decryption uses the secret key, preshared with the PC during the device's enrollment phase, to decrypt each block back to plaintext. After decrypting all blocks, PKCS\#7 padding is removed from the final block to recover the original ECG signal. 

\begin{figure*}[!htbp]
\centering    
\includegraphics[width=0.97\linewidth]{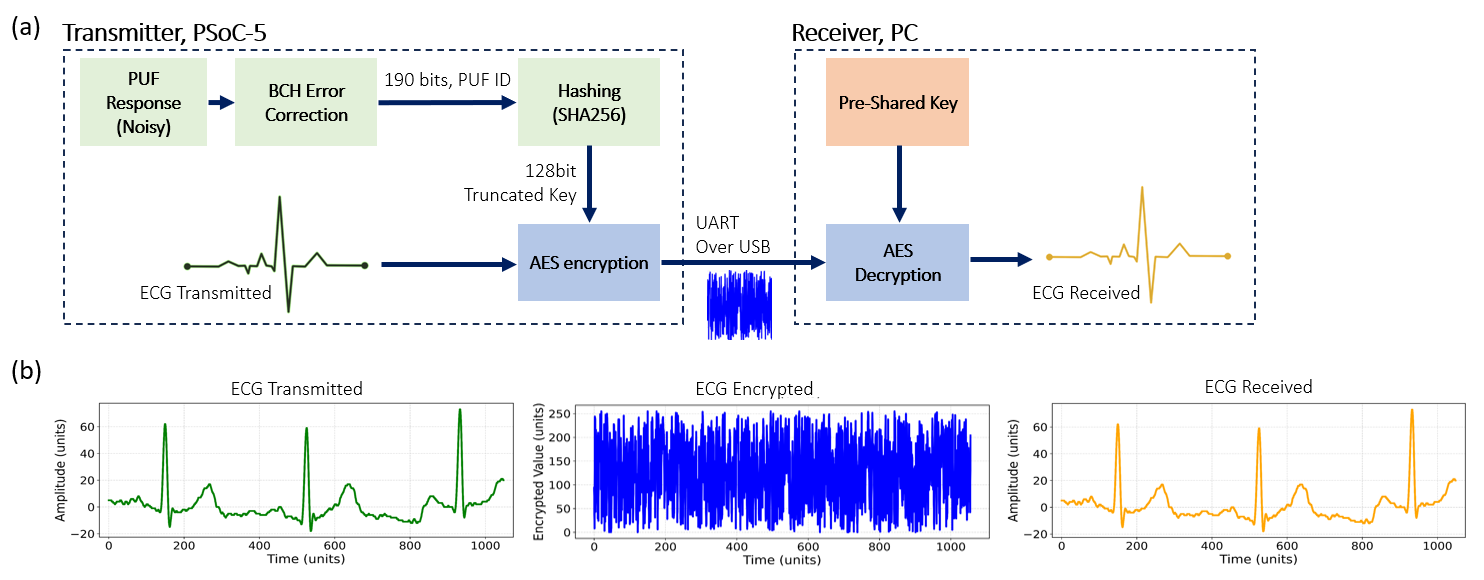}
\caption{(a) Schematic of the encryption and decryption datapath for data transfers from the PSoC-5 to a connected PC using the PUF-derived secret key. 
(b) Plots of ECG data showing the original signal, the AES-128–encrypted signal transmitted over the channel, and the decrypted signal recovered at the PC.}
\label{fig_secure_communication_demo}
\end{figure*}

Figure~\ref{fig_secure_communication_demo}(a) presents the schematic of the datapath. Figure~\ref{fig_secure_communication_demo}(b) shows the ECG waveform stored on the PSoC-5, transmitted in encrypted form, and subsequently received and decrypted on the PC. The decrypted waveform is identical to the original hardcoded ECG signal, demonstrating the working of the PUF-based secret key generation and AES encryption framework. Additionally, the encrypted signal appears random and exhibits no visible characteristics of the original ECG waveform, confirming that the encryption successfully hides the underlying ECG data during transmission.

\subsection{Resource and Energy Usage Characterization}

Secret key generation and encryption rely on several security components operating together: the PUF circuit, the response-generation algorithm, BCH error correction, SHA-256 hashing, and AES-128 encryption. All of these modules run on the PSoC-5, which provides only 256~KB of Flash and 64~KB of SRAM; resources that must also accommodate the application’s functional code. Given these constraints, it is essential to quantify how much memory each security component consumes and whether the entire pipeline can be integrated without exceeding device limits. Beyond memory usage, the latency, power consumption, and overall energy cost of the security component determine whether the design is suitable for low-power, real-time embedded applications. To this end, we characterized the resource usage and energy consumption of the security components. The measurements were conducted across multiple project configurations on the PSoC-5, each enabling a different subset of the security modules. Table~\ref{PUF_Characterisation} summarizes the results.

Regarding Flash usage, the results show that the PUF circuit and Algorithm~\ref{alg:ioPUF} occupy 3.1\% of the total available memory. Adding BCH increases the usage to 6\%, while integrating the SHA-256 hashing library raises it to 6.4\%. The complete system, comprising the PUF circuit, Algorithm~\ref{alg:ioPUF}, BCH, SHA-256, and AES, utilizes only 7.7\% of the total Flash memory, corresponding to 19.8~KB.

In terms of latency, the system exhibits a maximum delay of 600~ms for the configuration that includes all the security components. Latency remains relatively low for intermediate configurations, such as 45~ms for the PUF circuit alone and 150~ms when BCH encoding is added.

For power consumption, the PUF circuit draws the highest average power at 84.9~mW. Adding BCH and cryptographic functions slightly reduces the average power, with the full ECG encryption application consuming 79~mW. However, the total energy consumption increases as more functionality is incorporated, rising from 3.8~mJ for the basic PUF to 47.4~mJ for the complete system. This increase is primarily due to longer execution times introduced by additional operations.

Overall, the Flash consumption of only a few tens of kilobytes, millisecond-scale latencies, and milliwatt-level power usage indicate that these components impose lightweight resource overheads, demonstrating their suitability for deployment in embedded devices with constrained memory and energy resources.

\textit{Note:} The Flash usage of  project configuration was determined using the PSoC Creator IDE~\cite{psocCreatorIDE}, which reports the amount of program memory utilized by the firmware. Latency was measured by toggling a dedicated I/O pin at the start and end of program execution. The resulting pulse width was observed on an oscilloscope to determine the execution time. For power consumption measurements, a digital multimeter was connected in series with the PSoC-5 device to record 1000 current readings, which were averaged to obtain $I_{\text{avg}}$. The average power, $P_{\text{avg}}$, was then calculated using $P_{\text{avg}} = V \times I_{\text{avg}}$, where $V$ is the supply voltage. Finally, the average energy per operation, $E_{\text{avg}}$, was computed as $E_{\text{avg}} = P_{\text{avg}} \times t$, where $t$ is the duration of a single execution cycle for each project configuration.

\begin{table*}[!htbp]
\centering
\caption{Flash usage, latency, power, and energy consumption of different ioPUF+ project configurations.}
\label{PUF_Characterisation}
    \resizebox{0.90\textwidth}{!}{%
    \begin{tabular}{|p{3.5cm}|c|c|c|c|}
        \hline
        \textbf{Project Configuration} & \textbf{Flash (KB)} & \textbf{Latency (ms)} & \textbf{Avg. Power (mW)} & \textbf{Avg. Energy (mJ)} \\
        \hline

        PUF Alone & 8 (3.1\%) & 45 & 84.9 & 3.8 \\
        \hline
        PUF + BCH & 15.3 (6\%) & 150 & 80.5 & 12.8 \\
        \hline
        PUF + BCH + Hash & 16.4 (6.4\%) & 175 & 82 & 14.3 \\
        \hline
        PUF + BCH + Hash + AES  & 19.8 (7.7\%) & 600 & 79 & 47.4 \\
        \hline
    \end{tabular}
    }
\end{table*}

%% file: myContents/section-6-conclusion.tex
\section{Conclusion}
\label{sec_conclusion}

In this work, we present ioPUF+, which introduces a novel PUF that generates unique fingerprints for Integrated Circuits (ICs) utilizing pull-up and pull-down resistor measurements on the I/O pins of these ICs. The method is based on the observation that manufacturing process variations cause the resistor values to differ from their nominal design values in a manner that is unique to each IC.
The current PUF design is an extension of our earlier work, ioPUF~\cite{ioPUF}; compared to that, the PUF proposed in this work generates responses of length fourfold higher, i.e., from 45 to 190 bits, and improved performance metrics. We implemented the PUF and characterized it on 32 PSoC-5 microcontrollers from Infineon. The work reports a detailed characterization of PUF responses, including stability analysis of the responses under varying temperature and voltage conditions. The results indicate good PUF performance. Beyond introducing the PUF, ioPUF+ also details the method and implementation of converting PUF responses to derive secret keys using Bose–Chaudhuri–Hocquenghem (BCH) error correction and Secure Hash Algorithm (SHA-256). Finally, an application of ioPUF+ secret keys is demonstrated by securing device-to-device data transfers using AES encryption. A detailed analysis of the resource and power consumption of the complete ioPUF+ system is also reported. The full implementation requires only 19.8 KB of Flash, exhibits a latency of 600 ms, and consumes 79 mW of power, demonstrating the suitability of ioPUF+ for resource-constrained IoT nodes.

Future work includes evaluating the feasibility of deploying ioPUF+ on additional microcontroller families, investigating alternative error-correction schemes and hardware accelerators, and exploring PUF-based mechanisms for lightweight asymmetric key generation on ultra-low-power IoT platforms. We also intend to study the long-term aging effects of I/O resistor networks and their impact on the PUF responses.

\section*{Acknowledgment}
The authors used generative AI tools for grammar, language refinement, and stylistic editing during manuscript preparation. All technical content, analyses, and conclusions were conceived, validated, and written by the authors.